\documentstyle[epsf,axodraw]{article}
\begin{document}

\bigskip
{\Large\bf\centerline{Double Scattering Effect in Transverse Momentum Distribution} 
\centerline{of Inclusive $J/\psi$ Production in Photo-Nucleus Collision}}
\bigskip
\normalsize

\centerline{Han-Wen Huang\footnote{Email:huanghw@physik.uni-bielefeld.de}}
\centerline{Fakult\"{a}t f\"{u}r Physik, Universit\"{a}t Bielefeld}
\centerline{D-33501 Bielefeld, Germany} 
\bigskip

\begin{abstract}
In terms of multiple scattering picture, we calculate the double scattering
effect in the transverse momentum distribution of $J/\psi$ photoproduction. 
Applying the generalized factorization theorem, we find that the contributions
from double scattering can be expressed in terms of twist-4 nuclear parton 
correlation functions, which is the same as that used to explain the nuclear
dependence in di-jet momentum imbalance and in direct photon production.
Using the known information on the twist-4 parton correlation functions,
we estimate that the double scattering contributes a small suppression in 
$J/\psi$ photoproduction. In the analysis we only take into account the 
leading order in the
small velocity expansion for the nonperturbative parts related to the 
quarkonium.  
\end{abstract}

\vspace{1cm}
PACS number(s): 12.38.-t, 12.38.Bx, 11.80.La

Keywords: High twist, Photoproduction, $J/\psi$

\vfill\eject

\section{Introduction}

\hspace{0.2in} Heavy quarkonium production in high-energy collisions affords us 
a variety of insights into the underlying dynamics of the strong interaction.
Especially in nucleus-nucleus collisions, we may find that a suppression
of $J/\psi$ production can serve as a clear signature of the existence
of the deconfining phase of QCD \cite{satz}. This suppression effect was
observed by NA38 collaboration later \cite{na38}. However, successive
observation \cite{e772} pointed out that in proton-nucleus collisions, in 
which there is no QGP formed, such suppression could also exist. There have 
been a number of attempts to explain it, such as pre-resonance absorption, 
gluon shadowing,
hadronic co-mover absorption and the energy loss model 
\cite{15,qiu,wong,5,6,7}, etc. To understand the $J/\psi$ suppression mechanism
clearly, its production by scattering in different nuclear facilities must be
studied carefully.

In principle, the $J/\psi$ state can be described in a Fock state
decomposition
\begin{equation}
|J/\psi>=O(1)|c\bar{c}(^3S^{(1)}_1)>+O(v)|c\bar{c}(^3P^{(8)}_J)g>+
\cdots,
\end{equation}
where $^{2S+1}L^{(1,8)}_J$ characterizes the quantum state of the
$c\bar{c}$ pair with color-singlet (C1) or color-octet (C8) respectively. 
This expression is
valid for the non-relativistic QCD(NRQCD) framework and the coefficients
of each component depend on $v$, the small relative velocity between quark
and antiquark. The charmonium production can be divided into two steps.
The first step is the production of $c\bar{c}$ pair. The $c\bar{c}$ pair
can be either $(c\bar{c})_1$ or $(c\bar{c})_8$, and are produced
perturbatively and almost instantaneously, with a formation time
$\tau_f\approx(2m_c)^{-1}=0.07fm$ in the $c\bar{c}$ rest frame. The
second step is the formation of a physical state of $J/\psi$ from $c\bar{c}$ 
pair, that need
longer time. The second step is a nonperturbative process, and in
different environments such as vacuum, hadron matter and QGP, the
original produced $c\bar{c}$ pairs undergo different transition rules to
final observable $J/\psi$. All these processes have been studied carefully
before. However, besides this nonperturbative effect related to the
transitions there is another nonperturbative effect if there are hadrons
or nucleus in the initial states. This nonperturbative effect can be
analyzed with a twist expansion method \cite{ellis,sterman}. Ref. \cite{ma} 
gives the effect of the order of next-to-leading twist in inclusive
photoproduction of quarkonium. In this paper we will use factorization
at higher twist to describe multiple scattering effect for $J/\psi$
production in photo-nucleus collision, and our result can be extended
to study the corresponding cases of hardon-nucleus and nucleus-nucleus 
collisions.

In terms of factorization at higher twist, Luo, Qiu and Sterman (LQS)
\cite{luo,luo1} have developed a consistent treatment of multiple scattering
at partonic level. They showed that nuclear effect may be brought
directly into the scattering formalism of QCD, by treating it as a
factorizable nonleading-order correction to hard scattering, and nuclear
enhancement appears in this context as a property of multi-parton matrix
elements. In \cite{luo} they derive the anomalous nuclear dependence of
jet cross sections in elastic scattering and photoproduction in terms of
twist-four parton distributions in nuclei. In \cite{luo2} they expressed
the nuclear dependence of di-jet moment imbalance in photo-nucleus
collisions and estimated the size of the relevant twist-4 parton
distributions by using the Fermi Lab E683 data. Recently Guo \cite{guo}
applied this method to Drell-Yan process, and calculated its transverse
distribution in hadron-nucleus collisions.

All above processes are totally inclusive, and the extension to our case
is not quite straightforward. The process we consider is seminclusive,
where a $c\bar{c}$ pair is observed indirectly. We neglect the higher order
$v^2$ contributions and consider that the formation of $J/\psi$ is totally 
evoluted from the leading order $c\bar{c}(^3S^{(1)}_1)$ pair. This formation 
starts with a C1 or C8 $c\bar{c}$ pair produced from initial hard 
collision between
photon and a parton from nucleus, and then on its way out of the
nucleus, it undergoes a soft scattering with another parton from the
nucleus to form a color single $c\bar{c}(^3S^{(1)}_1)$ pair. The soft
scattering for $c\bar{c}$ pair can be done by either its quark or its 
antiquark picking up the
soft gluon from nucleus. Our paper is divided into four parts. In section II, we
derived the general formalism and the result of single scattering. In
section III, the double scattering contributions are derived. In the
last section, we present the numerical results and give a brief
discussion.

\section{General formalism and the single scattering result}

\hspace{0.2in} To study the nuclear effect in transverse momentum distribution, we 
consider the differential cross section for the inclusive process:
\begin{equation}
\gamma(p^{\prime})+A(p)\rightarrow J/\psi(Q)+X.
\end{equation}
The momenta of particles are given in brackets, $p^{\prime}$ is the momentum
of incoming photon, $p$ is the momentum per nucleon for nuclear target,
and $Q$ is the momentum of outgoing $J/\psi$. In terms 
of contributions from multiple scattering, we expand the differential cross
section as
\begin{equation}
\frac{d\sigma_{\gamma A}}{dQ^2_{\perp}dy}=\frac{d\sigma^S_{\gamma A}}
{dQ^2_{\perp}dy}+\frac{d\sigma^D_{\gamma A}}{dQ^2_{\perp}dy}+\cdots
\end{equation}
The superscript ``S" and ``D" denote contributions from the ``single
scattering" and ``double scattering", respectively, and ``$\cdots$"
represents contributions from even higher multiple scattering. We choose a frame in which
the nucleus $A$ moves in the $z$-direction and the photon in the opposite 
direction, $Q_{\perp}$ is the transverse momentum of the $J/\psi$, and $y$ is
its rapidity
\begin{equation}
y=\frac{1}{2}ln\frac{Q_0+Q_z}{Q_0-Q_z}.
\end{equation}
In our approximation the quarkonium mass $m_{\psi}$ is twice of quark
mass $m_c$.
We define quantities $s=(p+p^{\prime})^2$, $t=(p-Q)^2$ and 
$u=(p^{\prime}-Q)^2$.  For performing the analysis it
is convenient to use light-cone coordinate system. In this system the photon
carries the momentum $p^{\prime\mu}=(0,p^{\prime -},0_{\perp})$, and
$p^{\mu}=(p^+,0,0_{\perp})$. We introduce in this frame two vectors $n$ and
$\bar{n}$ and a tensor $d^{\mu\nu}_{\perp}$:
\begin{eqnarray}\nonumber
n^{\mu}&=&(0,1,0_{\perp}),~~\bar{n}^{\mu}=(1,0,0_{\perp}),\\
d^{\mu\nu}_{\perp}&=&g^{\mu\nu}-n^{\mu}\bar{n}^{\nu}
-n^{\nu}\bar{n}^{\mu}.
\end{eqnarray}
We will work in the light-cone gauge $n\cdot A(x)=A^+(x)=0$, where
$A^{\mu}(x)=A^{a,\mu}(x)T^a$ is the gluon field.

In order to compare with experimental data, we calculate the ratio of
total differential cross section and single scattering contribution,
\begin{equation}
\rho=\frac{d\sigma_{\gamma A}}{dQ^2_{\perp}dy}\left/
\frac{d\sigma^S_{\gamma A}}{dQ^2_{\perp}dy}\approx
1+\frac{d\sigma^D_{\gamma A}}{dQ^2_{\perp}dy}\right/
\frac{d\sigma^S_{\gamma A}}{dQ^2_{\perp}dy}.
\end{equation}

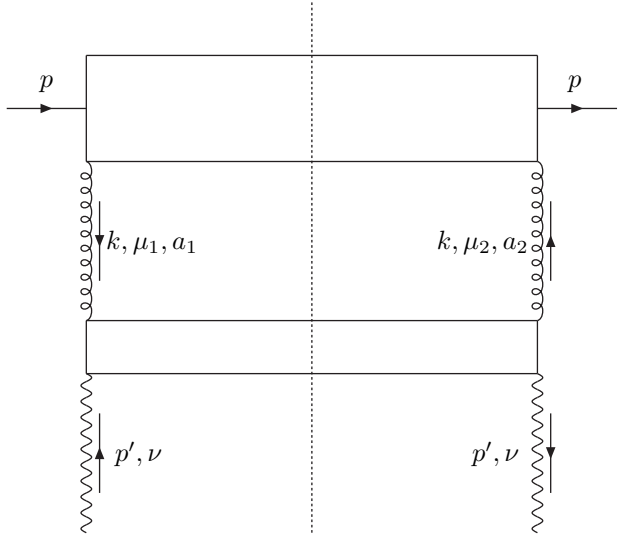
\begin{figure}\label{fig1}
\begin{center}\begin{picture}(250,300)(0,0)
\ArrowLine(10,200)(40,200)\Text(25,210)[]{$p$}
\Line(40,180)(40,220)\Line(210,180)(210,220)
\Line(40,220)(210,220)\Line(40,180)(210,180)
\ArrowLine(210,200)(240,200)\Text(225,210)[]{$p$}
\Line(40,120)(210,120)\Line(40,100)(210,100)
\Line(40,100)(40,120)\Line(210,100)(210,120)
\Gluon(40,180)(40,120){2}{10}\ArrowLine(45,165)(45,135)
\Text(65,150)[]{$k,\mu_1,a_1$}
\Gluon(210,180)(210,120){2}{10}\ArrowLine(215,135)(215,165)
\Text(190,150)[]{$k,\mu_2,a_2$}
\Photon(40,40)(40,100){2}{10}\ArrowLine(45,55)(45,85)
\Text(60,70)[]{$p^{\prime},\nu$}
\Photon(210,40)(210,100){2}{10}\ArrowLine(215,85)(215,55)
\Text(195,70)[]{$p^{\prime},\nu$}
\DashLine(125,40)(125,240){1.0}
\end{picture}

\caption{A Graphical representation for single scattering process in 
photon-nucleus collision.} 
\end{center}
\end{figure}

To analyze the effect of single scattering we need to consider the diagram 
in Fig.1, in which the up box represents the nonperturbative part related
to the
nucleus A, the down box contains the perturbative part for production of the
$c\bar{c}$ pair at the leading order of coupling constant and the transition of
the pair into $J/\psi$, they are illustrated in the Feynman diagram given in 
Fig.2. The 
contribution from Fig.1 can be written as:
\begin{equation}
2s\sigma^S_{\gamma A}=\int\frac{d^4k}{(2\pi)^4}
\hat{S}^{a_1a_2}_{\mu_1\mu_2}(k)\int d^4ze^{ik\cdot z}
<p_A|A^{a_1\mu_1}(z)A^{a_2\mu_2}(0)|p_A>,
\end{equation}
where $\hat{S}^{a_1a_2}_{\mu_1\mu_2}(k)$ corresponds to the lower part in
Fig.1, its leading order part is shown in Fig.2, where the left and
right box include all possible tree Feynman diagrams with the external
partons shown in the figures. In the center of mass frame of high energy 
collision, all partons inside the nucleus are parallel to each other, along 
the momentum of the nucleus. In order to get the leading order contribution, 
we make the collinear expansion,
\begin{eqnarray}\nonumber
\hat{S}^{a_1a_2}_{\mu_1\mu_2}(k)&=&\left.\hat{S}^{a_1a_2}_{\mu_1\mu_2}(xp)+
\frac{\partial \hat{S}^{a_1a_2}_{\mu_1\mu_2}(k)}{\partial k^{\alpha}}\right|
_{k=xp}\Delta k^{\alpha}\\
&+&\left.\frac{1}{2}\frac{\partial^2\hat{S}^{a_1a_2}_{\mu_1\mu_2}(k)}{\partial
k^{\alpha}\partial k^{\beta}}\right|_{k=xp}\Delta k^{\alpha}\Delta k^{\beta}
+\cdots,
\end{eqnarray}
Where $\Delta k=k-xp$. We only keep the first term 
$\hat{S}^{a_1a_2}_{\mu_1\mu_2}(xp)$, which is
independent of $k^-$ and $k_{\perp}$, so we can perform the integration
over $k^-$ and $k_{\perp}$ directly,
\begin{eqnarray}\nonumber
2s\sigma^S_{\gamma A}&=&\int\frac{dx}{x}\hat{S}^{a_1a_2}_{\mu_1\mu_2}(xp)
\frac{xp^+}{2\pi}\int
dz^-e^{ixp^+z^-}<p_A|A^{a_1\mu_1}(z)A^{a_2\mu_2}(0)|p_A>\\
&=&\int\frac{dx}{x}f_{g/A}(x)\left(-\frac{1}{16}d_{\perp}^{\mu_1\mu_2}\delta^{a_1a_2}
\hat{S}^{a_1a_2}_{\mu_1\mu_2}(xp)\right),  
\end{eqnarray}
where we have used the matrix element relation\cite{ma}
\begin{equation}
<p_A|A^{a_1\mu_1}(z)A^{a_2\mu_2}(0)|p_A>=\frac{1}{16}
d_{\perp}^{\mu_1\mu_2}\delta^{a_1a_2}
<p_A|A^{a\perp}(z)A^a_{\perp}(0)|p_A>.
\end{equation}
The function $f_{g/A}(x)$ is defined as
\begin{equation}\label{function}
f_{g/A}(x)=-\frac{xp^+}{2\pi}\int
dz^-e^{ixp^+z^-}<p_A|A^{a\perp}(z)A^a_{\perp}(0)|p_A>.
\end{equation}

\begin{figure}
\begin{center}\begin{picture}(120,150)(0,0)
\Gluon(5,140)(20,110){1}{8}\Text(5,145)[]{$\mu_1,a_1$}
\Photon(5,10)(20,40){1}{8}\Text(10,10)[]{$\nu$}
\Gluon(115,140)(100,110){1}{8}\Text(115,145)[]{$\mu_2,a_2$}
\Photon(115,10)(100,40){1}{8}\Text(110,10)[]{$\nu$}
\Line(20,110)(20,40)\Line(20,40)(30,40)\Line(30,40)(30,110)
\Line(30,110)(20,110)
\Line(100,110)(100,40)\Line(100,40)(90,40)\Line(90,40)(90,110)
\Line(90,110)(100,110)
\Gluon(30,40)(90,40){1}{10}
\Text(45,50)[]{$\lambda$}\Text(75,50)[]{$\lambda$}
\ArrowLine(30,110)(50,85)\ArrowLine(50,85)(30,60)
\ArrowLine(50,85)(70,85)
\ArrowLine(70,85)(90,110)\ArrowLine(90,60)(70,85)
\DashLine(60,140)(60,10){1.0}
\end{picture}

\caption{The diagram corresponds to the lower part in Fig.1 for $J/\psi$
production at lowest order partonic level.}
\end{center}
\end{figure}
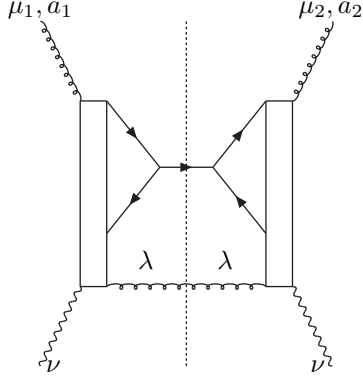

We consider the formation of $J/\psi$ from a $c\bar{c}$ pair at leading
order in $v^2$. According to the NRQCD factorization formalism\cite{braaten}, 
the amplitude of $J/\psi$ production from partons can
be factorized into

\begin{equation}
{\cal A}(g\gamma\rightarrow \psi_cg)={\cal A}(g\gamma\rightarrow
c\bar{c}(^3S_1^{(1)}))\left|_{pert.}
\otimes{\cal A}(c\bar{c}(^3S_1^{(1)})\rightarrow
\psi_c)\right|_{nopert.}.
\end{equation}
Here we only keep the leading order contribution comes from color singlet 
$c\bar{c}(^3S^{(1)}_1)$ pair. Contribution from color octet and other higher angular 
momentum $c\bar{c}$ pair states are suppressed at least $v^2$, we neglect them
here. $c\bar{c}(^3S^{(1)}_1)$ pair production starts at $O(\alpha_s^2)$ via process 
$\gamma g\rightarrow\psi g$, whose Feynman diagram is shown in Fig.3. 
Now $\hat{S}^{a_1a_2}_{\mu_1\mu_2}(xp)$ can be written as
\begin{equation}\label{frac}
\hat{S}^{a_1a_2}_{\mu_1\mu_2}(xp)=\frac{1}{2}\Gamma^{(2)}\times L^{a_1}_{\mu_1\nu\lambda}(k)
\otimes R^{a_2}_{\lambda\nu\mu_2}(k)\times{\cal M}(\psi_c),
\end{equation}
where $\frac{1}{2}$ represents the average over the initial photon spin, 
\begin{equation}
\Gamma^{(2)}=\frac{dQ^2_{\perp}dy}{8\pi}\frac{1}{s+t-m^2_{\psi}}
\delta(x+\frac{u}{s+t-m^2_{\psi}})
\end{equation}
represents the final two body phase space integration for $J/\psi$ and a
gluon. Nonperturbative part ${\cal M}(\psi_c)$ in Eq.(\ref{frac}) cannot be computed from first
principles without resorting to lattice QCD, we neglect the low energy
dependence of interactions, the nonperturbative amplitude only affect the
overall normalization constant, 
\begin{equation}\label{nonper}
{\cal M}(\psi_c)=\frac{1}{6m_c}<0|O^{(1)}_{\psi}(^3S_1)|0>=constant.
\end{equation}

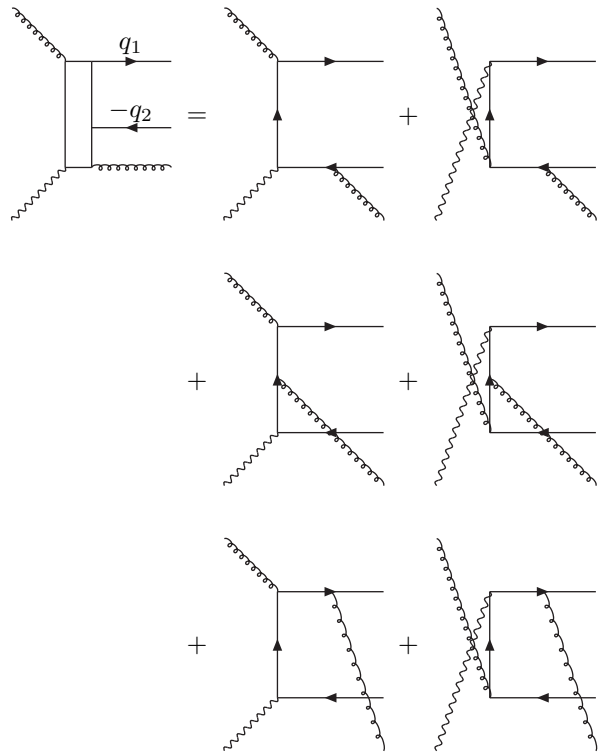
\begin{figure}
\begin{center}\begin{picture}(240,300)(0,0)
\Gluon(10,290)(30,270){1}{8}\Photon(10,210)(30,230){1}{8}
\Line(30,270)(30,230)\Line(30,230)(40,230)\Line(40,230)(40,270)
\Line(40,270)(30,270)
\ArrowLine(40,270)(70,270)\ArrowLine(70,245)(40,245)
\Gluon(40,230)(70,230){1}{8}
\Text(80,250)[]{=}
\Text(55,277)[]{$q_1$}\Text(55,252)[]{$-q_2$}

\Gluon(90,290)(110,270){1}{8}\Photon(90,210)(110,230){1}{8}
\ArrowLine(150,230)(110,230)\ArrowLine(110,230)(110,270)
\ArrowLine(110,270)(150,270)
\Gluon(130,230)(150,210){1}{8}
\Text(160,250)[]{+}

\Gluon(170,290)(190,230){1}{15}\Photon(170,210)(190,270){1}{15}
\ArrowLine(230,230)(190,230)\ArrowLine(190,230)(190,270)
\ArrowLine(190,270)(230,270)
\Gluon(210,230)(230,210){1}{8}
\Text(80,150)[]{+}

\Gluon(90,190)(110,170){1}{8}\Photon(90,110)(110,130){1}{8}
\ArrowLine(150,130)(110,130)\ArrowLine(110,130)(110,170)
\ArrowLine(110,170)(150,170)
\Gluon(110,150)(150,110){1}{15}
\Text(160,150)[]{+}

\Gluon(170,190)(190,130){1}{15}\Photon(170,110)(190,170){1}{15}
\ArrowLine(230,130)(190,130)\ArrowLine(190,130)(190,170)
\ArrowLine(190,170)(230,170)
\Gluon(190,150)(230,110){1}{15}
\Text(80,50)[]{+}

\Gluon(90,90)(110,70){1}{8}\Photon(90,10)(110,30){1}{8}
\ArrowLine(150,30)(110,30)\ArrowLine(110,30)(110,70)
\ArrowLine(110,70)(150,70)
\Gluon(130,70)(150,10){1}{10}
\Text(160,50)[]{+}

\Gluon(170,90)(190,30){1}{15}\Photon(170,10)(190,70){1}{15}
\ArrowLine(230,30)(190,30)\ArrowLine(190,30)(190,70)
\ArrowLine(190,70)(230,70)
\Gluon(210,70)(230,10){1}{10}

\end{picture}

\caption{Diagrams represent the $c\bar{c}$ pair production amplitude including
all partonic hard subprocess at leading order in perturbative coupling 
constant.}
\end{center}
\end{figure}

In (\ref{frac}) $L^{a_1}_{\mu_1\nu\lambda}$ and $R^{a_2}_{\lambda\nu\mu_2}$
represent the perturbative production amplitude the of $c\bar{c}(^3S^{(1)}_1)$
pair shown in the left and right parts of Fig.2 respectively. We have
\begin{eqnarray}\label{defi}\nonumber
L^{a_1}_{\mu_1\nu\lambda}(k)&=&\frac{1}{\sqrt{3}}\sum_{s_1,s_2}
\bar{u}(q_1;s_1)\bar{L}^{a_1}_{\mu_1\nu\lambda}(k,q_1,q_2)v(q_2,s_2)
<\frac{1}{2}s_1;\frac{1}{2}s_2|1S_z>\\
&=&\frac{1}{\sqrt{3}}Tr\left[\bar{L}^{a_1}_{\mu_1\nu\lambda}(k,q_1,q_2)
P_{1S_z}(q_1,q_2)\right],
\end{eqnarray}
where $\bar{L}^{a_1}_{\mu_1\nu\lambda}(k,q_1,q_2)$ represents the left box
of Fig.2 with external quark momentum $q_1$ and antiquark momentum $q_2$, and 
its all tree level diagrams are illustrated in
the diagrams of Fig.3. The heavy quark and antiquark propagate nearly on
-shell with a large combined $Q=q_1+q_2$ and a small relative momentum
$\bar{q}=q_1-q_2$. The spin quantum numbers for the $c\bar{c}$ pair are
projected
out by the sums over $SU(2)$ Clebsch-Gordan coefficients
$<\frac{1}{2}s_1;\frac{1}{2}s_2|1S_z>$. The $4\times 4$ matrix is
\begin{eqnarray}\nonumber
P_{1S_z}&=&\sum_{s_1s_2}v(q_2;s_2)\bar{u}(q_1;s_1)<\frac{1}{2}s_1;
\frac{1}{2}s_2|SS_z>\\
&=&\frac{-1}{2\sqrt{2}m_c}(\rlap/{q_1}-m_c)\rlap/{\epsilon}(Q,S_z)
(\rlap/{q_2}+m_c)
\end{eqnarray}
up to $O(q^2)$ corrections. The project operators' dependence on $\bar{q}$
is irrelevant for $S$-wave $c\bar{c}$ production. This indicates 
$q_1=q_2=q$, and $Q=2q$, so 
\begin{equation}
P_{1S_z}=\frac{-1}{2\sqrt{2}m_c}
(\rlap/{q}-m_c)\rlap/{\epsilon}(Q,S_z)(\rlap/{q}+m_c).
\end{equation}
The final perturbative result can be derived as\cite{ma}
\begin{eqnarray}\label{leading}
\frac{d\sigma^S_{\gamma A}}{dQ^2_{\perp}dy}&=&\frac{32\pi\alpha\alpha^2_s e^2_Q 
m_{\psi}|R(0)|^2}{3s(s+t-m_{\psi}^2)}\frac{f_{g/A}(x)}{x}
\left\{\frac{\hat{s}^2(\hat{s}-m^2_{\psi})^2+\hat{t}^2(\hat{t}^2-m^2_{\psi})^2
+(\hat{s}+\hat{t})^2(\hat{s}+\hat{t}-m^2_{\psi})^2}
{(\hat{s}-m^2_{\psi})^2(\hat{t}-m^2_{\psi})^2(\hat{s}+\hat{t})^2}\right\},
\end{eqnarray}
where 
\begin{equation}
\hat{s}=(xp+p^{\prime})^2,~~\hat{t}=(xp-Q)^2,
\end{equation}
and $x$ is fixed according to the momentum conservation,
\begin{equation}
x=-\frac{u}{s+t-m_{\psi}^2}.
\end{equation}

\section{The double scattering result}

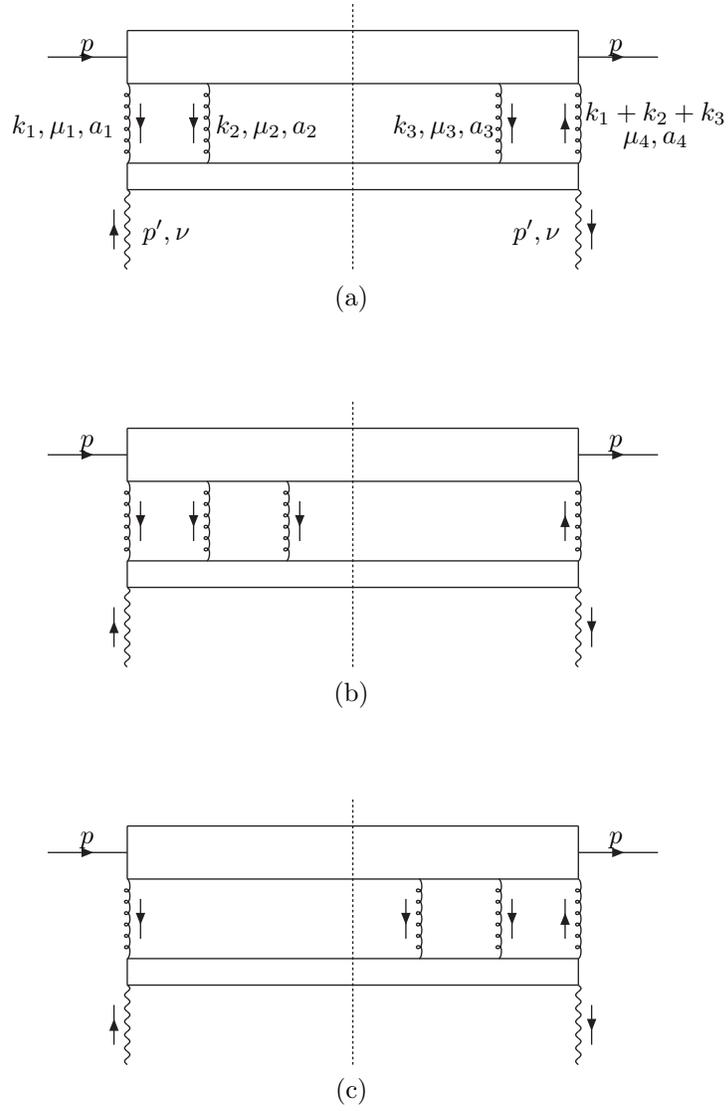
\begin{figure}

\begin{center}\begin{picture}(260,450)(0,0)
\ArrowLine(10,400)(40,400)\Text(25,405)[]{$p$}
\Line(40,390)(40,410)\Line(210,390)(210,410)
\Line(40,410)(210,410)\Line(40,390)(210,390)
\ArrowLine(210,400)(240,400)\Text(225,405)[]{$p$}
\Line(40,360)(210,360)\Line(40,350)(210,350)
\Line(40,350)(40,360)\Line(210,350)(210,360)
\Gluon(40,390)(40,360){1}{6}\ArrowLine(45,382.5)(45,367.5)
\Gluon(210,390)(210,360){1}{6}\ArrowLine(205,367.5)(205,382.5)
\Gluon(70,390)(70,360){1}{6}\ArrowLine(65,382.5)(65,367.5)
\Gluon(180,390)(180,360){1}{6}\ArrowLine(185,382.5)(185,367.5)
\Photon(40,320)(40,350){1}{6}\ArrowLine(35,327.5)(35,342.5)
\Photon(210,320)(210,350){1}{6}\ArrowLine(215,342.5)(215,327.5)
\DashLine(125,320)(125,420){1.0}
\Text(93,375)[]{$k_2,\mu_2,a_2$}\Text(16,375)[]{$k_1,\mu_1,a_1$}
\Text(240,380)[]{$k_1+k_2+k_3$}\Text(240,370)[]{$\mu_4,a_4$}
\Text(160,375)[]{$k_3,\mu_3,a_3$}\Text(55,335)[]{$p^{\prime},\nu$}
\Text(195,335)[]{$p^{\prime},\nu$}
\Text(125,310)[]{(a)}

\ArrowLine(10,250)(40,250)\Text(25,255)[]{$p$}
\Line(40,240)(40,260)\Line(210,240)(210,260)
\Line(40,260)(210,260)\Line(40,240)(210,240)
\ArrowLine(210,250)(240,250)\Text(225,255)[]{$p$}
\Line(40,210)(210,210)\Line(40,200)(210,200)
\Line(40,200)(40,210)\Line(210,200)(210,210)
\Gluon(40,240)(40,210){1}{6}\ArrowLine(45,232.5)(45,217.5)
\Gluon(210,240)(210,210){1}{6}\ArrowLine(205,217.5)(205,232.5)
\Gluon(70,240)(70,210){1}{6}\ArrowLine(65,232.5)(65,217.5)
\Gluon(100,240)(100,210){1}{6}\ArrowLine(105,232.5)(105,217.5)
\Photon(40,170)(40,200){1}{6}\ArrowLine(35,177.5)(35,192.5)
\Photon(210,170)(210,200){1}{6}\ArrowLine(215,192.5)(215,177.5)
\DashLine(125,170)(125,270){1.0}
\Text(125,160)[]{(b)}

\ArrowLine(10,100)(40,100)\Text(25,105)[]{$p$}
\Line(40,90)(40,110)\Line(210,90)(210,110)
\Line(40,110)(210,110)\Line(40,90)(210,90)
\ArrowLine(210,100)(240,100)\Text(225,105)[]{$p$}
\Line(40,60)(210,60)\Line(40,50)(210,50)
\Line(40,50)(40,60)\Line(210,50)(210,60)
\Gluon(40,90)(40,60){1}{6}\ArrowLine(45,82.5)(45,67.5)
\Gluon(210,90)(210,60){1}{6}\ArrowLine(205,67.5)(205,82.5)
\Gluon(150,90)(150,60){1}{6}\ArrowLine(145,82.5)(145,67.5)
\Gluon(180,90)(180,60){1}{6}\ArrowLine(185,82.5)(185,67.5)
\Photon(40,20)(40,50){1}{6}\ArrowLine(35,27.5)(35,42.5)
\Photon(210,20)(210,50){1}{6}\ArrowLine(215,42.5)(215,27.5)
\DashLine(125,20)(125,120){1.0}
\Text(125,10)[]{(c)}

\end{picture}


\caption{Graphical representation of double scattering contribution in
photon-nucleus collisions; a) real diagram, b) and c) interference
diagram.}

\end{center}

\end{figure}

\hspace{0.2in}Now we begin to consider the double scattering effect, at lowest order,
the cross section corresponding to soft rescattering are shown in Fig.4, 
its contribution can be written as
\begin{eqnarray}\label{double}
\nonumber
2sd\sigma^D_{\gamma A}&=&\int\frac{d^4k_1d^4k_2d^4k_3}{(2\pi)^4(2\pi)^4(2\pi)^4}
\hat{C}^{a_1a_2a_3a_4}_{\mu_1\mu_2\mu_3\mu_4}(k_1,k_2,k_3)\int d^4z_1d^4z_2
d^4z_3e^{ik_1\cdot z_1+ik_2\cdot z_2+k_3\cdot z_3}\\
&&<p_A\left|A^{a_1\mu_1}(z_1)A^{a_2\mu_2}(z_2)A^{a_3\mu_3}(z_3)
A^{a_4\mu_4}(0)\right|p_A>.
\end{eqnarray}
The momenta, color- and Lorentz- indices are marked in Fig.4. In Eq.(\ref{double})
$\hat{C}^{a_1a_2a_3a_4}_{\mu_1\mu_2\mu_3\mu_4}(k_1,k_2,k_3)$ corresponds to hard
production of $c\bar{c}$ pair and its transition to $J/\psi$. To pick up
the leading contributions to the nuclear enhancement, we expand the parton
momenta at the following according their collinear values:
\begin{eqnarray}\nonumber
k_1&=&x_1p^+\bar{n},\\\nonumber
k_2&=&x_2p^+\bar{n}+k_{2\perp},
\\k_3&=&x_3p^+\bar{n}+k_{3\perp}.
\end{eqnarray}
The minus components of the $k_i$
give even smaller contributions in
$\hat{C}^{a_1a_2a_3a_4}_{\mu_1\mu_2\mu_3\mu_4}(k_1,k_2,k_3)$ and will be neglected.
In addition, $k_{1\perp}$ dependence in $\hat{C}^{a_1a_2a_3a_4}_{\mu_1\mu_2\mu_3\mu_4}$ will not associated with
$A$ enhancement and we drop it as well. Later, we will show that $x_2$ and $x_3$
can be fixed by poles as functions of $k_{2\perp}$ (or $k_{3\perp}$) and
$Q_{\perp}$,
and that they vanish when $k_{2\perp}$ and $k_{3\perp}$ go to zero.

Once we have dropped the $k_{1\perp}$ and $k_i^-$ (i=1,2,3) dependence in
$\hat{C}^{a_1a_2a_3a_4}_{\mu_1\mu_2\mu_3\mu_4}$, their integrals give $\delta$ functions and allow us rewrite 
Eq.(\ref{double}) as
\begin{eqnarray}\nonumber
2sd\sigma^D_{\gamma A}&=&\int\frac{p^+dx_1}{2\pi}\int\frac{p^+dx_2}{2\pi}\frac
{d^2k_{2\perp}}{(2\pi)^2}\int\frac{p^+dx_3}{2\pi}\frac{d^2k_{3\perp}}{(2\pi)^2}
\int dz_1^-dz_2^-d^2z_{2\perp}dz_3^-d^2z_{3\perp}\\\nonumber
&&e^{ix_1p^+z^-_1+ix_2p^+z_2^-+ix_3p^+z^-_3}
e^{ik_{2\perp}\cdot z_{2\perp}+ik_{3\perp}\cdot z_{3\perp}}\\
&&\hat{C}^{a_1a_2a_3a_4}_{\mu_1\mu_2\mu_3\mu_4}(k_1,k_2,k_3)
<p_A|A^{a_1\mu_1}(z_1)
A^{a_2\mu_2}(z_2)A^{a_3\mu_3}(z_3)A^{a_4\mu_4}(0)|p_A>.
\end{eqnarray}
First of all, we expand the hard part $C^{a_1a_2a_3a_4}_{\mu_1\mu_2\mu_3\mu_4}(k_1,k_2,k_3)$ 
in momenta about the collinear direction:
\begin{eqnarray}\label{expan}\nonumber
&&\hat{C}^{a_1a_2a_3a_4}_{\mu_1\mu_2\mu_3\mu_4}(k_1,k_2,k_3)\hspace{2mm}
=\hspace{2mm}\hat{C}^{a_1a_2a_3a_4}_{\mu_1\mu_2\mu_3\mu_4}(x_1p,
x_2p,x_3p)\\\nonumber
&+&\left.\left.\frac{\partial\hat{C}^{a_1a_2a_3a_4}_{\mu_1\mu_2\mu_3\mu_4}
(x_1p,k_2,x_3p}
{\partial k_{2\perp}^{\rho}}\right|_{k_{2\perp}=0_{\perp}}k_{2\perp}^{\rho}+
\frac{\partial\hat{C}^{a_1a_2a_3a_4}_{\mu_1\mu_2\mu_3\mu_4}(x_1p,x_2p,k_3)}
{\partial k_{3\perp}^{\rho}}\right|_{k_{3\perp}=0_{\perp}}k_{3\perp}^{\rho}\\
&+&\left.\frac{\partial^2\hat{C}^{a_1a_2a_3a_4}_{\mu_1\mu_2\mu_3\mu_4}(x_1p,k_2,k_3)}
{2\partial k_{2\perp}^{\rho}\partial k_{3\perp}^{\sigma}}\right|_{k_{2\perp}=k_{3\perp}=0_{\perp}}
k_{2\perp}^{\rho}k_{3\perp}^{\sigma}+\cdots.
\end{eqnarray}
In this discussion we will confine ourselves to experiments with unpolarized
beams.
Thus odd-twist contributions of the second and third terms in Eq.(\ref{expan})
vanish on taking a spin average. In addition, the leading term 
$C^{a_1a_2a_3a_4}_{\mu_1\mu_2\mu_3\mu_4}(x_1p,x_2p,x_3p)$
does not contribution to $A$ dependence because it is independent of
transverse momenta. We did not list explicitly terms such as 
$\partial^2\hat{C}/\partial k^2_{2\rho}$ and 
$\partial^2\hat{C}/\partial k^3_{2\sigma}$ because, for reasons that will become
clear below, they do not contribute physical double scattering.
Therefore, double scattering effect comes entirely from
the last single term of Eq.(\ref{expan})
\begin{eqnarray}\nonumber
2sd\sigma^D_{\gamma A}&=&\left.\frac{1}{2}\int\frac{(p^+)^3dx_1dx_2dx_3}{(2\pi)^3}\int dz_1^-dz_2^-dz_3^-\left(\frac{\partial^2 
\hat{C}^{a_1a_2a_3a_4}_{\mu_1\mu_2\mu_3\mu_4}(x_1p,k_2,k_3)}
{\partial k_2^{\rho}\partial k_3^{\sigma}}\right|_{k_{2\perp}=k_{3\perp}=0_{\perp}}
\right)\\\nonumber 
&&e^{ix_1p^+z^-_1+ix_2p^+z^-_2+ix_3p^+z^-_3}\\\nonumber
&&\int\frac{d^2k_{2\perp}}{(2\pi)^2}\frac{d^2k_{3\perp}}{(2\pi)^2}
k_{2\perp}^{\rho}k_{3\perp}^{\sigma}
e^{ik_{2\perp}\cdot z_{2\perp}+ik_{3\perp}\cdot z_{3\perp}}
<p_A|A^{\mu_1}_{a_1}(z_1)
A^{\mu_2}_{a_2}(z_2)A^{\mu_3}_{a_3}(z_3)A^{\mu_4}_{a_4}(0)|p_A>\\\nonumber
&=&\left.-\frac{1}{2}\int\frac{(p^+)^3dx_1dx_2dx_3}{(2\pi)^3}\int dz_1^-dz_2^-dz_3^-
\left(\frac{\partial^2\hat{C}^{a_1a_2a_3a_4}_{\mu_1\mu_2\mu_3\mu_4}(x_1p,k_2,k_3)}
{\partial k_{2\perp}^{\rho}\partial k_{3\perp}^{\sigma}}\right|_{k_{2\perp}=k_{3\perp}=0_{\perp}}
\right)\\
&&e^{ix_1p^+z^-_1+ix_2p^+z^-_2+ix_3p^+z^-_3}
\left(<p_A\left| A^{\mu_1}_{a_1}(z^-_1)\frac{\partial A^{\mu_2}_{a_2}(z_2)}
{\partial z_{2\rho}}\frac{\partial A^{\mu_3}_{a_3}(z_3)}
{\partial z_{3\sigma}}A^{\mu_4}_{a_4}(0)\right|p_A>\right).
\end{eqnarray}
Because the values of $x_1,x_2$ and $x_3$ will be fixed by $\delta$ functions
and poles of the hard part $\hat{C}^{a_1a_2a_3a_4}_{\mu_1\mu_2\mu_3\mu_4}$, we
move the exponentials $e^{ix_1p^+z^-_1+ix_2p^+z^-_2+ix_3p^+z^-_3}$ and 
integrals inside the derivative of 
$\partial^2/\partial k^{\rho}_2\partial k_3^{\sigma}$,
\begin{eqnarray}\label{partia}
2sd\sigma^D_{\gamma A}&=&-\frac{1}{2}\int\frac{(p^+)^3dz_1^-dz_2^-dz_3^-}{(2\pi)^3}
\left(<p_A\left| A^{\mu_1}_{a_1}(z^-_1)\frac{\partial A^{\mu_2}_{a_2}(z_2)}
{\partial z_{2\rho}}\frac{\partial A^{\mu_3}_{a_3}(z_3)}
{\partial z_{3\sigma}}A^{\mu_4}_{a_4}(0)\right|p_A>\right)\\
&&
\left.\left(\frac{\partial^2}{\partial k_{2\perp}^{\rho}\partial 
k_{3\perp}^{\sigma}}\int dx_1dx_2dx_3
e^{ix_1p^+z^-_1+ix_2p^+z^-_2+ix_3p^+z^-_3} 
\hat{C}^{a_1a_2a_3a_4}_{\mu_1\mu_2\mu_3\mu_4}(x_1p,k_2,k_3)\right)
\right|_{k_{2\perp}=k_{3\perp}=0_{\perp}}.
\end{eqnarray}
According to the discussion in \cite{ma,luo}, the dominant contribution of
the four-gluon matrix element can be written as
\begin{eqnarray}\nonumber
&&<p_A\left| A^{\mu_1}_{a_1}(z^-_1)G^{\mu_2\rho}_{a_2}
(z^-_2)G^{\mu_3\sigma}_{a_3}(z^-_3)A^{\mu_4}_{a_4}(0)\right|p_A>
\\&=&
\frac{1}{4\times 8\times  8}\frac{p^{\mu_2}p^{\mu_3}}{(p^+)^2}
d^{\mu_1\mu_4}_{\perp}d^{\rho\sigma}_{\perp}
\delta^{a_1a_4}\delta^{a_2a_3}<p_A\left| A^{a\perp}(z^-_1)G_{+\perp}^b
(z^-_2)G^{b+\perp}(z^-_3)A_{\perp}^{a}(0)\right| p_A>
\end{eqnarray}
Therefore, the cross section for double scattering turns out to be
\begin{equation}\label{first}
d\sigma^D_{\gamma A}=\frac{1}{2s}(-\frac{1}{8^3})\int\frac{p^+dz^-_1dz^-_2dz^-_3}
{(2\pi)^3}
<p_A\left|A^{a\perp}(z^-_1)G_{+\perp}^b
(z^-_2)G^{b+\perp}(z^-_3)A_{\perp}^{a}(0)\right|p_A>
F(z^-_1,z^-_2,z^-_3),
\end{equation}
where
\begin{eqnarray}\label{second}\nonumber
F(z^-_1,z^-_2,z^-_3)&=&\frac{\partial^2}{\partial k_{2\perp\rho}\partial 
k_{3\perp}^{\rho}}
\left(\int dx_1dx_2dx_3e^{ix_1p^+z^-_1+ix_2p^+z^-_2+ix_3p^+z^-_3}\right.\\
&\times&\left.\left.\hat{C}^{a_1a_2a_3a_4}_{\mu_1\mu_2\mu_3\mu_4}(x_1p,k_2,k_3)p^{\mu_2}p^{\mu_3}
d^{\mu_1\mu_4}_{\perp}\delta_{a_2a_3}\delta_{a_1a_4}\right)
\right|_{k_{2\perp}=k_{3\perp}=0_{\perp}}.
\end{eqnarray}

Equation (\ref{first}) has been derived at lowest order in the hard
scattering. As it stands, it is not gauge invariant. Gauge invariant is
incorporated at higher orders and leading power by taking into account diagrams
involving more fields $n\cdot A$, which will form ordered exponentials between
the physical fields shown in (\ref{first}), as discussed, for instance, in
\cite{sterman,luo}. Such fields, however, can not correspond to physical
rescattering, since they can be eliminated by a change of gauge. This is the
reason why we neglected two derivatives with respect to $k_2$ and none with
respect to $k_3$, or vice versa. Such a term would involve only three physical
fields instead of four. As we shall see below, four physical fields are
required to produce nuclear effect.

From Eq.(\ref{first}) and (\ref{second}) all integrals of $x_1,x_2,x_3$ can
now be done explicitly without knowing the detail of the multi-parton
matrix elements. Now we are ready to discuss the calculation of the hard part
$\hat{C}^{a_1a_2a_3a_4}_{\mu_1\mu_2\mu_3\mu_4}(x_1p,k_2,k_3)$.
As is shown in the diagram of Fig.4, during the collision, firstly the photon
strongly interacts with a gluon from one nucleon in nucleus. After a hard
scattering, the outgoing $c\bar{c}$ pair interacts with a soft gluon from
another nucleon in the nucleus to form a color-singlet $c\bar{c}(^3S_1)$
pair. So there are two partons from nucleus participate in the scattering 
process. 
The kinematics can only fix one parton momentum, thus we need to integrate 
over the other parton's momentum. In
$\hat{C}^{a_1a_2a_3a_4}_{\mu_1\mu_2\mu_3\mu_4}(x_1p,k_2,k_3)$ we will 
encounter
one phase space $\delta$ function and two virtual propagator poles involving
$x_i's$, which correspond to the zero momenta fraction partons. The leading contribution of the integration over the extra parton 
momentum is given by the residues at these poles, which will fix $x_2$ and 
$x_3$ to be functions of the transverse momentum $k_{i\perp}$. Then in the 
limit $k_{i\perp}\rightarrow 0$ the $x_2$ and $x_3$ will vanish. Therefore 
$e^{ix_2p^+z^-_2}$ and $e^{ix_3p^+z^-_3}$ will reduce to unity eventually. 
We may expect a
significant contribution from the free integrals $\int dz^-_2dz^-_3$. This 
will be the origin of the $A$ dependence. 
  
Consider subprocess shown in Fig.4, there are three four-momentum linking the
partonic part and corresponding four gluon matrix element. In inclusive 
DIS \cite{luo} and Drell-Yan \cite{guo} processes, the final results depend 
only on the real diagram with one extra gluon in each side of the cut, 
and the role of interference diagrams with both gluons in one side of the cut 
is to take care of the infrared sensitivities of the short distance hard 
parts. The process we are considering is seminclusive and the subsequent 
$c\bar{c}$ pair is in color-singlet state. This demands that for the real 
scattering subprocess shown in Fig.4(a), the original $c\bar{c}$ pair produced 
from hard scattering of photon and one parton should be in color-octet state, 
while for the virtual subprocesses shown in Fig.4(b) and Fig.4(c),
the corresponding $c\bar{c}$ pair should be in color-singlet state. Thus we 
must consider the two different types of physical rescattering separately, 
and as we shall see below, each of their contributions remains infrared finite
in itself. This is quite different from the inclusive situations, in which 
divergences are canceled only after summing over all sorts of cut in a
Feynman diagram.

First of all, we consider contribution of real double scattering diagram 
illustrated in Fig.4(a). By the similar approach which we have taken in 
deriving Eq.(\ref{frac}), we can write its hard part 
$\hat{C}^{a_1a_2a_3a_4(1)}_{\mu_1\mu_2\mu_3\mu_4}$ as
\begin{equation}\label{cvalue}
\hat{C}^{a_1a_2a_3a_4(1)}_{\mu_1\mu_2\mu_3\mu_4}(x_1p,k_2,k_3)p^{\mu_2}p^{\mu_3}
=\frac{1}{2}\Gamma^{(2)}\times L^{a_1a_2}_{\mu_1\nu\lambda}(x_1p,k_2,k_3)\otimes
R^{a_3a_4}_{\lambda\nu\mu_4}(x_1p,k_2,k_3)
\times{\cal M}(\psi_c),
\end{equation}
Nonperturbative part ${\cal M}(\psi_c)$ has been defined in Eq.(\ref{nonper}).
The final state gluon quarkonium two particle phase space can be 
written as
\begin{equation}
\Gamma^{(2)}=\frac{dQ^2_{\perp}dy}{8\pi}\frac{1}{s+t-m^2_{\psi}}
\delta(x_1+x_2+\frac{u-2Q\cdot k_{2\perp}+k^2_{2\perp}}{s+t-m^2_{\psi}}).
\end{equation}

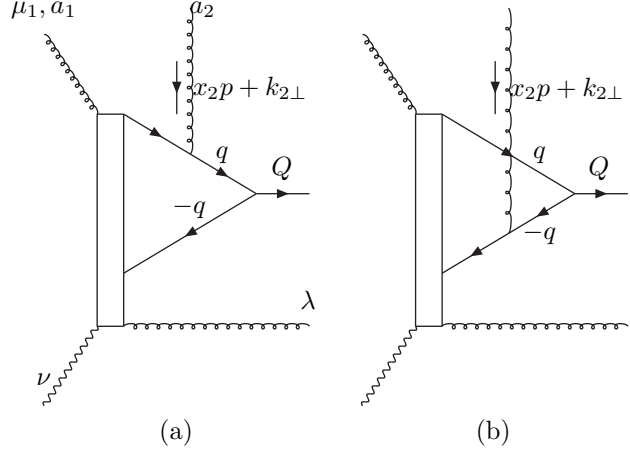
\begin{figure}
\begin{center}\begin{picture}(240,180)(0,0)
\Gluon(10,160)(30,130){1}{10}
\Text(10,170)[]{$\mu_1,a_1$}
\Photon(10,20)(30,50){1}{10}\Text(10,30)[]{$\nu$}
\Line(30,130)(30,50)\Line(30,50)(40,50)\Line(40,50)(40,130)
\Line(40,130)(30,130)
\ArrowLine(40,130)(65,115)\ArrowLine(65,115)(90,100)
\ArrowLine(90,100)(40,70)
\ArrowLine(90,100)(110,100)
\Gluon(65,170)(65,115){1}{10}\Text(70,170)[]{$a_2$}
\Gluon(40,50)(110,50){1}{15}\Text(110,60)[]{$\lambda$}
\ArrowLine(60,150)(60,130)
\Text(60,10)[]{(a)}
\Text(100,110)[]{$Q$}
\Text(77.5,115)[]{$q$}\Text(65,95)[]{$-q$}
\Text(88,140)[]{$x_2p+k_{2\perp}$}

\Gluon(130,160)(150,130){1}{10}\Photon(130,20)(150,50){1}{10}
\Line(150,130)(150,50)\Line(150,50)(160,50)\Line(160,50)(160,130)
\Line(160,130)(150,130)
\ArrowLine(160,130)(210,100)
\ArrowLine(210,100)(185,85)\ArrowLine(185,85)(160,70)
\ArrowLine(210,100)(230,100)
\Gluon(185,170)(185,85){1}{10}
\ArrowLine(180,150)(180,130)
\Gluon(160,50)(230,50){1}{15}
\Text(208,140)[]{$x_2p+k_{2\perp}$}
\Text(180,10)[]{(b)}
\Text(220,110)[]{$Q$}
\Text(197.5,85)[]{$-q$}
\Text(197.5,115)[]{$q$}

\end{picture}

\caption{Graphical representation of the left amplitude $L^{a_1a_2}
_{\mu_1\nu\lambda}$ in Eq.(\ref{cvalue}) for $c\bar{c}$ interacting with 
soft gluon, (a) 
shows the quark picks up the gluon, (b) shows the antiquark picks up the gluon}
\end{center}
\end{figure}

The represents of amplitudes $L^{a_1a_2}_{\mu_1\nu\lambda}$ and $R^{a_3a_4}_{\lambda\nu\mu_4}$ are shown in Fig.5 and Fig.6 respectively.
The $c\bar{c}$ pair can interacts with nucleon through one of its constituents 
quark or antiquark scattering with soft gluon. In case of the quark 
participating the interaction, as shown in Fig.5(a), we have
\begin{eqnarray}\label{la}\nonumber
L^{a_1a_2(a)}_{\mu_1\nu\lambda}(x_1p,k_2,k_3)&=&g_sTr\left\{\frac
{(\rlap/{q}-m_c)\rlap/{\epsilon}(Q,S_z)(\rlap/{q}+m_c)\rlap/{p}
(\rlap/{q}-x_2\rlap/{p}-\rlap/{k_{2\perp}}+m_c)}{2\sqrt{2}m_c[(q-x_2p-k_{2\perp})^2-m_c^2+i\varepsilon]}
T^{a_2}\bar{L}^{a_1}_{\mu_1\nu\lambda}(x_1p,q^{\prime},q)\right\}\\
&=&g_sTr\left\{\frac{(\rlap/{q}-m_c)\rlap/{\epsilon}(Q,S_z)(\rlap/{q}+m_c)
(2p\cdot q-\rlap/{p}\rlap/{k_{\perp}})T^{a_2}\bar{L}^{a_1}
_{\mu_1\nu\lambda}(x_1p,q^{\prime},q)}{2\sqrt{2}m_c
(k^2_{\perp}-2x_2p\cdot q-2q\cdot k_{\perp}+i\varepsilon)}\right\},
\end{eqnarray}
where $q^{\prime}=q-x_2p-k_{2\perp}$ is the momentum of outgoing quark after
hard scattering,
$\bar{L}^{a_1}_{\mu_1\nu\lambda}(x_1p,q^{\prime},q)$ has been defined in
Eq.(\ref{defi}) and shown in Fig.3. In Eq.(\ref{la}) we have taken 
$k_{2\perp}=-k_{3\perp}=k_{\perp}$ by momentum conservation.
In case of the antiquark participating the interaction, as shown in Fig.5(b), 
we have
\begin{eqnarray}\nonumber
L^{a_1a_2(b)}_{\mu_1\nu\lambda}(x_1p,k_2,k_3)&=&g_sTr\left\{\frac
{(-\rlap/{q}+x_2\rlap/{p}+\rlap/{k_{2\perp}}+m_c)\rlap/{p}(\rlap/{q}-m_c)
\rlap/{\epsilon}(Q,S_z)(\rlap/{q}+m_c)}
{2\sqrt{2}m_c[(-q+x_2p+k_{2\perp})^2-m_c^2+i\varepsilon]}
T^{a_2}\bar{L}^{a_1}_{\mu_1\nu\lambda}(x_1p,q,q^{\prime})\right\}\\
&=&g_sTr\left\{\frac{(-2p\cdot q+\rlap/{k_{\perp}}\rlap/{p})(\rlap/{q}-m_c)
\rlap/{\epsilon}(Q,S_z)(\rlap/{q}+m_c))T^{a_2}
\bar{L}^{a_1}_{\mu_1\nu\lambda}(x_1p,q,q^{\prime})}{2\sqrt{2}m_c
(k^2_{\perp}-2x_2p\cdot q-2q\cdot k_{\perp}+i\varepsilon)}\right\}.
\end{eqnarray}
The total contribution of Fig.5 is
\begin{eqnarray}\label{left}\nonumber
L^{a_1a_2}_{\mu_1\nu\lambda}(x_1p,k_2,k_3)&=&
L^{a_1a_2(a)}_{\mu_1\nu\lambda}(x_1p,k_2,k_3)
+L^{a_1a_2(b)}_{\mu_1\nu\lambda}(x_1p,k_2,k_3)\\
&=&\frac{-g_s}{2\sqrt{2}m_c}\frac{1}{x_2+\frac{2q\cdot k_{\perp}-k^2_{\perp}}{2p\cdot q}
-i\varepsilon}\tilde{L}^{a_1a_2}_{\mu_1\nu\lambda}(x_1p,k_2,k_3),
\end{eqnarray}
where
\begin{eqnarray}\label{left1}\nonumber
\tilde{L}^{a_1a_2}_{\mu_1\nu\lambda}(x_1p,k_2,k_3)&=&
Tr\left[(\rlap/{q}-m_c)
\rlap/{\epsilon}(Q,S_z)(\rlap/{q}+m_c)(1-\frac{\rlap/{p}\rlap/{k_{\perp}}}
{2p\cdot q})T^{a_2}\bar{L}^{a_1}_{\mu_1\nu\lambda}(x_1p,q^{\prime},q)\right.\\
&&\left.-\bar{L}^{a_1}_{\mu_1\nu\lambda}(x_1p,q,q^{\prime})T^{a_2}(1-
\frac{\rlap/{k_{\perp}}\rlap/{p}}{2p\cdot q})
(\rlap/{q}-m_c)\rlap/{\epsilon}(Q,S_z)(\rlap/{q}+m_c)\right].
\end{eqnarray}
In above Eq.(\ref{left}) we have explicitly extracted the pole contribution 
from the quark or antiquark propagator in amplitude
$L^{a_1a_2(a)}_{\mu_1\nu\lambda}(x_1p,k_2,k_3)$. 

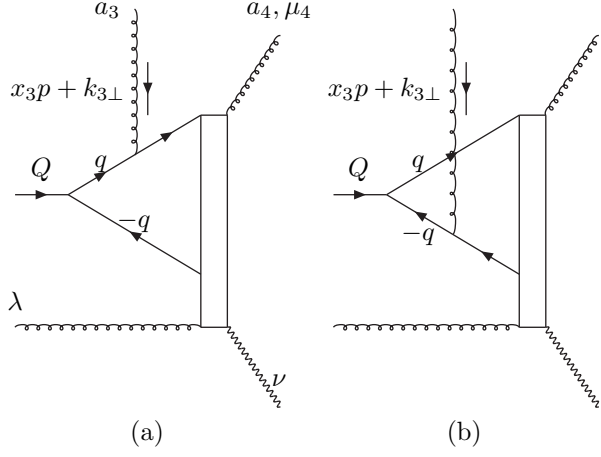
\begin{figure}
\begin{center}\begin{picture}(240,180)(0,0)
\ArrowLine(10,100)(30,100)
\ArrowLine(30,100)(55,115)\ArrowLine(55,115)(80,130)
\ArrowLine(80,70)(30,100)
\Line(80,130)(90,130)\Line(90,130)(90,50)
\Line(90,50)(80,50)\Line(80,50)(80,130)
\Gluon(10,50)(80,50){1}{15}\Text(10,60)[]{$\lambda$}
\Gluon(90,130)(110,160){1}{10}\Text(110,170)[]{$a_4,\mu_4$}
\Photon(90,50)(110,20){1}{15}\Text(110,30)[]{$\nu$}
\Gluon(55,170)(55,115){1}{10}
\Text(45,170)[]{$a_3$}\ArrowLine(60,150)(60,130)
\Text(60,10)[]{(a)}
\Text(20,110)[]{$Q$}
\Text(30,140)[]{$x_3p+k_{3\perp}$}
\Text(55,90)[]{$-q$}
\Text(42.5,112)[]{$q$}

\ArrowLine(130,100)(150,100)\ArrowLine(150,100)(200,130)
\ArrowLine(200,70)(175,85)\ArrowLine(175,85)(150,100)
\Line(200,130)(210,130)\Line(210,130)(210,50)
\Line(210,50)(200,50)\Line(200,50)(200,130)
\Gluon(130,50)(200,50){1}{15}
\Gluon(210,130)(230,160){1}{10}\Photon(210,50)(230,20){1}{15}
\Gluon(175,170)(175,85){1}{10}\ArrowLine(180,150)(180,130)
\Text(180,10)[]{(b)}
\Text(140,110)[]{$Q$}
\Text(150,140)[]{$x_3p+k_{3\perp}$}
\Text(162.5,85)[]{$-q$}
\Text(162.5,112)[]{$q$}

\end{picture}

\caption{Graphical representation for the right amplitude
$R^{a_3a_4}_{\lambda\nu\mu_4}$ in Eq.(\ref{cvalue}) 
for $c\bar{c}$ interacting with soft gluon, (a) 
shows the quark picks up the gluon, (b) shows the antiquark picks up the gluon}
\end{center}
\end{figure}

Similarly, we can derive the corresponding result for amplitude 
$R^{a_3a_4}_{\lambda\nu\mu_4}$ from the diagrams shown in Fig.6. 
Fig.6(a) represents the case of soft gluon attaching to heavy quark line, 
which gives
\begin{eqnarray}\nonumber
R^{a_3a_4(a)}_{\lambda\nu\mu_4}(x_1p,k_2,k_3)&=&g_sTr\left\{\frac
{(\rlap/{q}+x_3\rlap/{p}+\rlap/{k_{3\perp}}+m_c)\rlap/{p}(\rlap/{q}+m_c)
\rlap/{\epsilon}(Q,S_z)(\rlap/{q}-m_c)}
{2\sqrt{2}m_c[(q+x_3p+k_{3\perp})^2-m_c^2-i\varepsilon]}
T^{a_3}\bar{R}^{a_4}_{\lambda\nu\mu_4}(x_1p,q^{\prime\prime},q)\right\}\\
&=&g_sTr\left\{\frac{(2p\cdot q-\rlap/{k_{\perp}}\rlap/{p})(\rlap/{q}+m_c)
\rlap/{\epsilon}(Q,S_z)(\rlap/{q}-m_c)T^{a_2}
\bar{R}^{a_4}_{\lambda\nu\mu_4}(q^{\prime\prime},q)}{2\sqrt{2}m_c(
k^2_{\perp}+2x_3p\cdot q-2q\cdot k_{\perp}-i\varepsilon)}\right\},
\end{eqnarray}
where $q^{\prime\prime}=q+x_3p+k_{3\perp}$.
Fig.6(b) shows the case of soft gluon attaching to heavy antiquark, it
gives
\begin{eqnarray}\nonumber
R^{a_3a_4(b)}_{\lambda\nu\mu_4}(x_1p,k_2,k_3)&=&g_sTr\left\{\frac
{(\rlap/{q}+m_c)\rlap/{\epsilon}(Q,S_z)(\rlap/{q}-m_c)\rlap/{p}
(-\rlap/{q}-x_3\rlap/{p}-\rlap/{k_{3\perp}}+m_c)}{2\sqrt{2}m_c[(-q-x_3p-k_{3\perp})^2-m_c^2-i\varepsilon]}
T^{a_3}\bar{R}^{a_4}_{\lambda\nu\mu_4}(x_1p,q,q^{\prime\prime})\right\}\\
&=&g_sTr\left\{\frac{(\rlap/{q}+m_c)\rlap/{\epsilon}(Q,S_z)(\rlap/{q}+m_c)
(-2p\cdot q+\rlap/{p}\rlap/{k_{\perp}})T^{a_3}\bar{R}^{a_4}
_{\lambda\nu\mu_4}(x_1p,q,q^{\prime\prime})}{2\sqrt{2}m_c(
k^2_{\perp}+2x_3p\cdot q-2q\cdot k_{\perp}-i\varepsilon)}\right\}.
\end{eqnarray}
The total result of the amplitude $R^{a_3a_4}_{\lambda\nu\mu_4}$ is
\begin{eqnarray}\label{right}\nonumber
R^{a_3a_4}_{\lambda\nu\mu_4}(x_1p,k_2,k_3)&=&
R^{a_3a_4(a)}_{\lambda\nu\mu_4}(x_1p,k_2,k_3)
+R^{a_3a_4(b)}_{\lambda\nu\mu_4}(x_1p,k_2,k_3)\\
&=&\frac{g_s}{2\sqrt{2}m_c}\frac{1}{x_3-\frac{2q\cdot k_{\perp}-k^2_{\perp}}{2p\cdot q}
-i\varepsilon}\tilde{R}^{a_3a_4}_{\lambda\nu\mu_4}(x_1p,k_2,k_3),
\end{eqnarray}
where
\begin{eqnarray}\label{right1}\nonumber
\tilde{R}^{a_3a_4}_{\lambda\nu\mu_4}(x_1p,k_2,k_3)&=&
Tr\left[\bar{R}^{a_4}_{\lambda\nu\mu_4}
(x_1p,q^{\prime\prime},q)T^{a_3}(1-\frac{\rlap/{k_{\perp}}\rlap/{p}}{2p\cdot q})
(\rlap/{q}+m_c)\rlap/{\epsilon}(Q,S_z)(\rlap/{q}-m_c)\right.\\
&&\left.-(\rlap/{q}+m_c)
\rlap/{\epsilon}(Q,S_z)(\rlap/{q}-m_c)
(1-\frac{\rlap/{p}\rlap/{k_{\perp}}}
{2p\cdot q})T^{a_3}\bar{R}^{a_4}_{\lambda\nu\mu_4}(x_1p,q,q^{\prime\prime})\right].
\end{eqnarray}

Now we can perform the integration through $x_i(i=1,2,3)$ in function
$F(z^-_1,z^-_2,z^-_3)$. From Eq.(\ref{left}) and (\ref{right}), we see that both variables $x_2$ and $x_3$ have poles in
the upper half complex plane. So after fixing $x_1$ from $\delta$ function,
we can carry out the other integrals $\int dx_2dx_3$ by using contour
integrations in $x_2$ and $x_3$, closing at infinity, and circling the poles,
we get
\begin{eqnarray}\label{fvalue}\nonumber
F(z^-_1,z^-_2,z^-_3)&=&\frac{dQ^2_{\perp}dy}{16\pi}\frac{g^2_s}{s+t-m^2_{\psi}}
\frac{{\cal M}(\psi_c)}{8m^2_c}\frac{\partial^2}{\partial k^{\rho}_{\perp}
\partial k_{\perp\rho}}\left\{\int dx_1dx_2dx_3e^{ix_1p^+z^-_1+ix_2p^+z^-_2+ix_3p^+z^-_3}\right.\\\nonumber
&&\times\hspace{2mm}\frac{1}{(x_2+\frac{2q\cdot k_{\perp}-k^2_{\perp}}{2p\cdot q}
-i\varepsilon)}\frac{1}{(x_3-\frac{2q\cdot k_{\perp}+k^2_{\perp}}{2p\cdot q}
-i\varepsilon)}
\delta(x_1+x_2+\frac{u-2Q\cdot k_{\perp}-k^2_{\perp}}
{s+t-m^2_{\psi}})\\\nonumber
&&\left.\times\hspace{2mm}\tilde{L}^{a_1a_2}_{\mu_1\nu\lambda}(x_1p,k_2,k_3)
\otimes\tilde{R}^{a_3a_4}_{\lambda\nu\mu_4}(x_1p,k_2,k_3)d^{\mu_1\mu_4}_{\perp}
\delta^{a_1a_4}\delta^{a_2a_3}\right\}
_{k_{\perp}=0_{\perp}}\\\nonumber
&=&\frac{dQ^2_{\perp}dy}{16\pi}\frac{g^2_s}{s+t-m^2_{\psi}}
\frac{{\cal M}(\psi_c)}{8m^2_c}\frac{\partial^2}{\partial k^{\rho}_{\perp}
\partial k_{\perp\rho}}\left\{\int dx_2dx_3e^{ix_1p^+z^-_1+ix_2p^+(z^-_2-z^-_1)+ix_3p^+z^-_3}\right.\\\nonumber
&&\times\hspace{2mm}\frac{1}{(x_2+\frac{2q\cdot k_{\perp}-k^2_{\perp}}{2p\cdot q}
-i\varepsilon)}\frac{1}{(x_3-\frac{2q\cdot k_{\perp}-k^2_{\perp}}{2p\cdot q}
-i\varepsilon)}d^{\mu_1\mu_4}_{\perp}
\delta^{a_1a_4}\delta^{a_2a_3}\\\nonumber
&&\left.\times\hspace{2mm}\tilde{L}^{a_1a_2}_{\mu_1\nu\lambda}(x_1p,k_2,k_3)
\otimes\tilde{R}^{a_3a_4}_{\lambda\nu\mu_4}(x_1p,k_2,k_3)\right\}
_{k_{\perp}=0_{\perp}}\\\nonumber
&=&\frac{dQ^2_{\perp}dy}{16\pi}\frac{g^2_s}{s+t-m^2_{\psi}}
\frac{{\cal M}(\psi_c)}{8m^2_c}\frac{\partial^2}{\partial k^{\rho}_{\perp}
\partial k_{\perp\rho}}\left\{e^{ix_1p^+z^-_1+ix_2p^+(z^-_2-z^-_1)+ix_3p^+z^-_3}\right.\\\nonumber
&&\times\hspace{2mm}(i2\pi\theta(z^-_2-z^-_1))(i2\pi\theta(z^-_3))d^{\mu_1\mu_4}_{\perp}
\delta^{a_1a_4}\delta^{a_2a_3}\\
&&\times\hspace{2mm}\left.\left.\tilde{L}^{a_1a_2}_{\mu_1\nu\lambda}(x_1p,k_2,k_3)
\otimes\tilde{R}^{a_3a_4}_{\lambda\nu\mu_4}(x_1p,k_2,k_3)\right|_{
x_2=-x_3=-\frac{2q\cdot k_{\perp}-k^2_{\perp}}{2p\cdot q}}\right\}_{k_{\perp}=
0_{\perp}},
\end{eqnarray}
where
\begin{equation}
x_1=-\frac{u-2Q\cdot k_{\perp}-k_{\perp}^2}{s+t-m^2_{\psi}}.
\end{equation}
By substituting Eq.(\ref{fvalue}) into Eq.(\ref{first}), we can obtain the 
lowest order real double scattering cross section 
\begin{eqnarray}\label{diff}\nonumber
\frac{d\sigma^{D(1)}_{\gamma A}}{dQ^2_{\perp}dy}&=&-\hspace{1mm}\frac{g^2_s}{2s
\times 8^3}\frac{1}
{16\pi(s+t-m^2_{\psi})}\frac{{\cal M}(\psi_c)}{8m^2_c}
\int dz^-_1dz^-_2dz^-_3\frac{p^+}{2\pi}
\theta(z^-_z-z^-_1)\theta(z^-_3)\\\nonumber
&&<p_A\left| A^{\perp}_a(z^-_1)G_{+\perp}^b
(z^-_2)G^{+\perp}_b(z^-_3)A_{\perp}^{a}(0)\right|p_A>d^{\mu_1\mu_4}_{\perp}
\delta^{a_1a_4}\delta^{a_2a_3}\\\nonumber
&&\frac{\partial^2}{\partial k^{\rho}_{\perp}\partial k_{\perp\rho}}
\left\{e^{ix_1p^+z^-_1+ix_2p^+(z^-_2-z^-_1)+ix_3p^+z^-_3}
\tilde{L}^{a_1a_2}_{\mu_1\nu\lambda}(x_1p,k_2,k_3)\right.\\
&&\left.\left.\otimes\tilde{R}^{a_3a_4}_{\lambda\nu\mu_4}(x_1p,k_2,k_3)\right|_{
x_2=-x_3=-\frac{2q\cdot k_{\perp}-k^2_{\perp}}{2p\cdot q}}
\right\}_{k_{\perp}=0}.
\end{eqnarray}
One important step in getting the final result is taking the derivative
with respect to $k_{\perp}$ as define in Eq.(\ref{partia}). From 
Eq.(\ref{left1}) and (\ref{right1}),
it is obviously that when $k_{\perp}=0$, $x_2=x_3=0$, and $x_1
=x$, therefore $q^{\prime}=q^{\prime\prime}=q$, we found
\begin{equation}\label{cancel}
\tilde{L}^{a_1a_2}_{\mu_1\nu\lambda}(x_1p,k_2,k_3)|_{k_{\perp}=0}
=\tilde{R}^{a_3a_4}_{\lambda\nu\mu_4}(x_1p,k_2,k_3)|_{k_{\perp}=0}
=0.
\end{equation}
This indicates that there are cancellations in both the left and right 
amplitudes $\tilde{L}^{a_1a_2}_{\mu_1\nu\lambda}$ and 
$\tilde{R}^{a_3a_4}_{\lambda\nu\mu_4}$ of eq.(\ref{fvalue}). These are the cancellations between the
amplitudes of gluon interacting with heavy quark shown in Fig.5(a),6(a)
and corresponding ones of antiquark cases in Fig.5(b),6(b). These eliminate
the twist-two and even some twist-four terms contained in Eq.(\ref{double}).
Nevertheless certain twist four terms will survive and we study them in the
following. 

Eq.(\ref{cancel}) denotes that the derivatives on the exponential
$e^{ix_1p^+z^-_1+ix_2p^+(z^-_2-z^-_1)+ix_3p^+z^-_3}$ do not contribute,
and that we can therefore set
$e^{ix_1p^+z^-_1+ix_2p^+(z^-_2-z^-_1)+ix_3p^+z^-_3}=e^{ixp^+z^-_1}$. 
Now we defined four gluon matrix element
\begin{eqnarray}\label{distr}\nonumber
\frac{M_g(x)}{x}&=&\int\frac{p^+dz^-_1}{2\pi}dz^-_2\frac{dz^-_3}{2\pi}
e^{ixp^+z^-_1}\theta(z^-_z-z^-_1)\theta(z^-_3)\\
&&\times\hspace{2mm}<p_A\left| A^{a\perp}(z^-_1)G_{+\perp}^b
(z^-_2)G^{b+\perp}(z^-_3)A_{\perp}^{a}(0)\right|p_A>.
\end{eqnarray}
Substituting Eq.(\ref{distr}) into Eq.(\ref{diff}), we obtain the 
contribution from real double scattering as
\begin{eqnarray}\label{cross1}\nonumber
\frac{d\sigma^{D(1)}_{\gamma A}}{dQ^2_{\perp}dy}&=&-\hspace{2mm}
\frac{g^2_s}{2s\times 8^3}\frac{1}{8(s+t-m^2_{\psi})}\frac{M_g(x)}{x}
\frac{{\cal M}(\psi_c)}{8m^2_c}d^{\mu_1\mu_4}_{\perp}\delta^{a_1a_4}
\delta^{a_2a_3}\\
&&\left.\times\hspace{2mm}\frac{\partial^2}{\partial k^{\rho}_{\perp}
\partial k_{\perp\rho}}
\left\{\tilde{L}^{a_1a_2}_{\mu_1\nu\lambda}(x_1p,k_2,k_3)
\otimes\tilde{R}^{a_3a_4}_{\lambda\nu\mu_4}(x_1p,k_2,k_3)\right|_{
x_2=-x_3=-\frac{2q\cdot k_{\perp}-k^2_{\perp}}{2p\cdot q}}\right\}
_{k_{\perp}=0}.
\end{eqnarray}

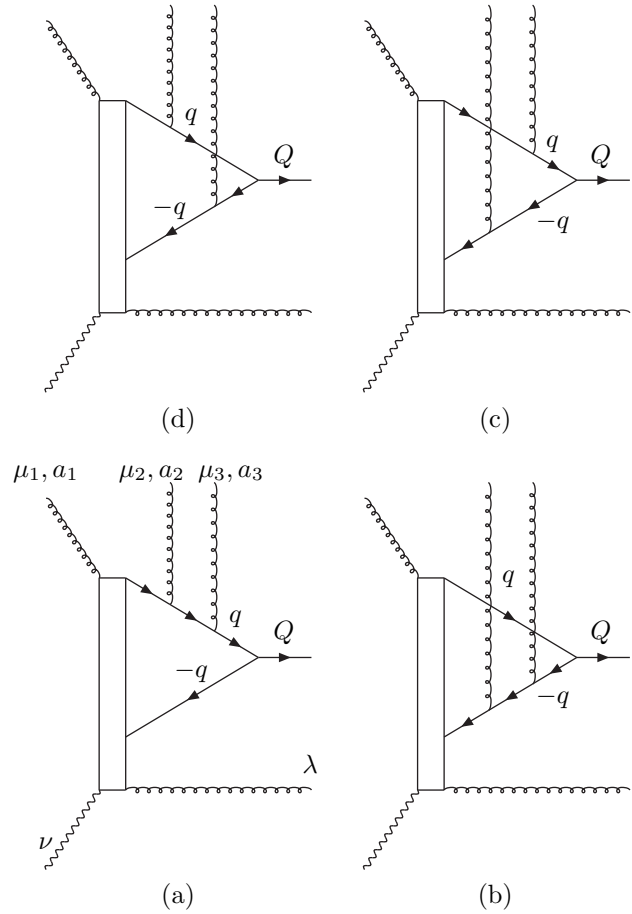
\begin{figure}
\begin{center}\begin{picture}(240,360)(0,0)
\Gluon(10,160)(30,130){1}{10}
\Text(10,170)[]{$\mu_1,a_1$}
\Photon(10,20)(30,50){1}{10}\Text(10,30)[]{$\nu$}
\Line(30,130)(30,50)\Line(30,50)(40,50)\Line(40,50)(40,130)
\Line(40,130)(30,130)
\ArrowLine(40,130)(56.67,120)\ArrowLine(56.67,120)(73.33,110)
\ArrowLine(73.33,110)(90,100)
\ArrowLine(90,100)(40,70)
\ArrowLine(90,100)(110,100)
\Gluon(56.67,166)(56.67,120){1}{10}\Text(50,170)[]{$\mu_2,a_2$}
\Gluon(73.33,166)(73.33,110){1}{10}\Text(80,170)[]{$\mu_3,a_3$}
\Gluon(40,50)(110,50){1}{15}\Text(110,60)[]{$\lambda$}
\Text(60,10)[]{(a)}
\Text(100,110)[]{$Q$}
\Text(81.67,115)[]{$q$}\Text(65,95)[]{$-q$}

\Gluon(130,160)(150,130){1}{10}\Photon(130,20)(150,50){1}{10}
\Line(150,130)(150,50)\Line(150,50)(160,50)\Line(160,50)(160,130)
\Line(160,130)(150,130)
\ArrowLine(160,130)(210,100)
\ArrowLine(210,100)(193.33,90)\ArrowLine(193.33,90)(176.67,80)
\ArrowLine(176.67,80)(160,70)
\ArrowLine(210,100)(230,100)
\Gluon(193.33,166)(193.33,90){1}{15}
\Gluon(176.67,166)(176.67,80){1}{15}
\Gluon(160,50)(230,50){1}{15}
\Text(180,10)[]{(b)}
\Text(220,110)[]{$Q$}
\Text(201.67,85)[]{$-q$}
\Text(185,130)[]{$q$}

\Gluon(10,340)(30,310){1}{10}
\Photon(10,200)(30,230){1}{10}
\Line(30,310)(30,230)\Line(30,230)(40,230)\Line(40,230)(40,310)
\Line(40,310)(30,310)
\ArrowLine(40,310)(90,280)
\ArrowLine(90,280)(73.33,270)\ArrowLine(73.33,270)(40,250)
\ArrowLine(90,280)(110,280)
\Gluon(56.67,346)(56.67,300){1}{10}
\Gluon(73.33,346)(73.33,270){1}{15}
\Gluon(40,230)(110,230){1}{15}
\Text(60,190)[]{(d)}
\Text(100,290)[]{$Q$}
\Text(65,305)[]{$q$}\Text(56.67,270)[]{$-q$}

\Gluon(130,340)(150,310){1}{10}\Photon(130,200)(150,230){1}{10}
\Line(150,310)(150,230)\Line(150,230)(160,230)\Line(160,230)(160,310)
\Line(160,310)(150,310)
\ArrowLine(160,310)(176.67,300)\Line(176.67,300)(193.33,290)
\ArrowLine(193.33,290)(210,280)
\ArrowLine(210,280)(176.67,260)
\ArrowLine(176.67,260)(160,250)
\ArrowLine(210,280)(230,280)
\Gluon(193.33,346)(193.33,290){1}{10}
\Gluon(176.67,346)(176.67,260){1}{15}
\Gluon(160,230)(230,230){1}{15}
\Text(180,190)[]{(c)}
\Text(220,290)[]{$Q$}
\Text(201.67,265)[]{$-q$}
\Text(201.67,295)[]{$q$}

\end{picture}

\caption{Graphical representation of the amplitude corresponding to the
left part of diagram in Fig.4(b); a) two soft gluons attach to quark
line, b) two soft gluons attach to antiquark line, c) and d) one gluon
attaches to quark line and another attaches to antiquark line}
\end{center}
\end{figure}

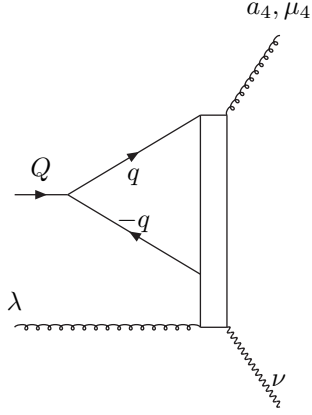
\begin{figure}
\begin{center}\begin{picture}(120,180)(0,0)
\ArrowLine(10,100)(30,100)
\ArrowLine(30,100)(80,130)
\ArrowLine(80,70)(30,100)
\Line(80,130)(90,130)\Line(90,130)(90,50)
\Line(90,50)(80,50)\Line(80,50)(80,130)
\Gluon(10,50)(80,50){1}{15}\Text(10,60)[]{$\lambda$}
\Gluon(90,130)(110,160){1}{10}\Text(110,170)[]{$a_4,\mu_4$}
\Photon(90,50)(110,20){1}{15}\Text(110,30)[]{$\nu$}
\Text(20,110)[]{$Q$}
\Text(55,90)[]{$-q$}
\Text(55,107)[]{$q$}

\end{picture}

\caption{Graphical representation of the amplitude corresponding to the right 
part of diagram in Fig.4(b)}
\end{center}
\end{figure}

Now we begin to consider the contribution from the virtual diagrams
shown in Fig.4(b) and Fig.4(c). The graphical represents of the amplitudes 
corresponding to the left and right parts of diagram in Fig.4(b) are given 
in Fig.7 and Fig.8 respectively, and the corresponding represents for Fig.4(c) 
can also be given similarly. Following the same methods as that we use 
to derive Eq.(\ref{cross1}), we obtain the
differential cross section coming from virtual double scattering as
\begin{eqnarray}\nonumber
\frac{d\sigma^{D(2)}_{\gamma A}}{dQ^2_{\perp}dy}&=&-\hspace{2mm}
\frac{g^2_s}{2s\times 8^3}\frac{1}{8(s+t-m^2_{\psi})}\frac{M_g(x)}{x}
\frac{{\cal M}(\psi_c)}{8m^2_c}d^{\mu_1\mu_4}_{\perp}\delta^{a_1a_4}
\delta^{a_2a_3}\\\nonumber
&&\left.\times\hspace{2mm}\frac{\partial^2}{\partial k^{\rho}_{\perp}
\partial k_{\perp\rho}}
\left\{\tilde{L}^{a_1a_2a_3}_{\mu_1\nu\lambda}(x_1p,k_2,k_3)
\otimes\tilde{R}^{a_4}_{\lambda\nu\mu_4}(x_1p,k_2,k_3)\right|_{
x_2=-\frac{2q\cdot k_{\perp}-k^2_{\perp}}{2p\cdot q},
x_3=\frac{2q\cdot k_{\perp}+k^2_{\perp}}{2p\cdot q}}\right.\\
&&\left.\left.+\hspace{2mm}\tilde{L}^{a_1}_{\mu_1\nu\lambda}(x_1p,k_2,k_3)
\otimes\tilde{R}^{a_2a_3a_4}_{\lambda\nu\mu_4}(x_1p,k_2,k_3)\right|_{
x_2=-\frac{2q\cdot k_{\perp}-k^2_{\perp}}{2p\cdot q},
x_3=\frac{2q\cdot k_{\perp}+k^2_{\perp}}{2p\cdot q}}\right\}
_{k_{\perp}=0},
\end{eqnarray}
where
\begin{eqnarray}\nonumber
\tilde{L}^{a_1a_2a_3}_{\mu_1\nu\lambda}(x_1p,k_2,k_3)&=&
Tr\left\{T^{a_2}T^{a_3}(1-\frac{\rlap/{k_{\perp}}\rlap/{p}}
{2p\cdot q})(\rlap/{q}-m_c)\rlap/{\epsilon}(Q,S_z)(\rlap/{q}+m_c)\right.\\
&&\left.(1+\frac{\rlap/{p}\rlap/{k_{\perp}}}
{2p\cdot q})\bar{L}^{a_1}_{\mu_1\nu\lambda}(x_1p,\bar{q}_1,\bar{q}_2)\right\},
\\\nonumber
\tilde{R}^{a_2a_3a_4}_{\lambda\nu\mu_4}(x_1p,k_2,k_3)&=&
Tr\left\{T^{a_3}T^{a_2}(1+\frac{\rlap/{k_{\perp}}\rlap/{p}}
{2p\cdot q})(\rlap/{q}+m_c)\rlap/{\epsilon}(Q,S_z)(\rlap/{q}-m_c)\right.\\
&&\left.(1-\frac{\rlap/{p}\rlap/{k_{\perp}}}
{2p\cdot q})\bar{R}^{a_4}_{\lambda\mu\mu_1}(x_1p,\bar{q}_1,\bar{q}_2)\right\},\\
\tilde{L}^{a_1}_{\mu_1\nu\lambda}(x_1p,k_2,k_3)&=&
Tr\left[(\rlap/{q}-m_c)\rlap/{\epsilon}(Q,S_z)(\rlap/{q}+m_c)
\bar{L}^{a_1}_{\mu_1\nu\lambda}(xp,q,q)\right]\\
\tilde{R}^{a_4}_{\lambda\nu\mu_4}(x_1p,k_2,k_3)&=&
Tr\left[(\rlap/{q}+m_c)
\rlap/{\epsilon}(Q,S_z)(\rlap/{q}-m_c)
\bar{R}^{a_4}_{\lambda\mu\mu_1}(xp,q,q)\right],
\end{eqnarray}
where
\begin{eqnarray}\nonumber
x_1&=&x-\frac{k_{\perp}^2}{q\cdot p},\\\nonumber
\bar{q}_1&=&q-x_2p-k_{\perp},\\\nonumber
\bar{q}_2&=&q-x_3p+k_{\perp}.
\end{eqnarray}

Now we can work out the derivative, the calculation is tedious but 
straightforward. The final result for lowest order double scattering cross
section is
\begin{eqnarray}\label{final}\nonumber
\frac{d\sigma^D_{\gamma A}}{dQ^2_{\perp}dy}&=&
\frac{d\sigma^{D(1)}_{\gamma A}}{dQ^2_{\perp}dy}+
\frac{d\sigma^{D(2)}_{\gamma A}}{dQ^2_{\perp}dy}\\
&=&\frac{16\pi^3\alpha\alpha^3_s e^2_Q 
|R(0)|^2}{s(s+t-m_{\psi}^2)m^3_{\psi}}\frac{M_{g/A}(x)}{x}
\frac{f(\hat{s},\hat{t},m^2_{\psi})}
{(\hat{s}-m^2_{\psi})^4(\hat{t}-m^2_{\psi})^4(\hat{s}+\hat{t})^4},
\end{eqnarray}
where the function $f$ is
\begin{eqnarray}\nonumber
f(\hat{s},\hat{t},m^2_{\psi})&=&\left\{-72Q_{\perp}^2m_{\psi}^{16}\hat{s}^2
+16Q_{\perp}^2m_{\psi}^{14}\hat{s}^2(82\hat{s}+73\hat{t})
+8Q_{\perp}^2m_{\psi}^{12}\hat{s}^2(-179\hat{s}^2-550\hat{s}\hat{t}
-88\hat{t}^2)\right.\\\nonumber
&+&8Q_{\perp}^2m_{\psi}^{10}\hat{s}^2(-89\hat{s}^3+518\hat{s}^2 
\hat{t}-135\hat{s}\hat{t}^2-158\hat{t}^3)\\\nonumber 
&+&8Q_{\perp}^2m_{\psi}^8\hat{s}^2(387\hat{s}^4+290\hat{s}^3\hat{t}
+1327\hat{s}^2 \hat{t}^2+1204\hat{s}\hat{t}^3+337\hat{t}^4)\\\nonumber
&+&8Q_{\perp}^2m_{\psi}^6\hat{s}^2(-402\hat{s}^5-870\hat{s}^4\hat{t}
-2263\hat{s}^3\hat{t}^2- 2560\hat{s}^2\hat{t}^3-1811\hat{s}\hat{t}^4
-516\hat{t}^5)\\\nonumber 
&+& 8Q_{\perp}^2m_{\psi}^4\hat{s}^2 (128 \hat{s}^6 + 498 \hat{s}^5 \hat{t} +
1529 \hat{s}^4 \hat{t}^2 + 2222 \hat{s}^3 \hat{t}^3 + 2182 \hat{s}^2 \hat{t}^4 + 1160 \hat{s} 
\hat{t}^5 + 306 \hat{t}^6)\\\nonumber 
&+& 256 Q_{\perp}^2 m_{\psi}^2 \hat{s}^3 \hat{t} ( - \hat{s}^5 - 3 \hat{s}^4 
\hat{t} - 5 \hat{s}^3 \hat{t}^2 - 5 \hat{s}^2 \hat{t}^3 - 3 \hat{s} 
\hat{t}^4 - \hat{t}^5)\\\nonumber 
&+&9m_{\psi}^{18}(\hat{s}^2+2\hat{s}\hat{t}+\hat{t}^2)
+4m_{\psi}^{16}(-53\hat{s}^3-23\hat{s}^2\hat{t} +
 17 \hat{s} \hat{t}^2 - 13 \hat{t}^3)\\\nonumber 
&+& m_{\psi}^{14} ( - 181 \hat{s}^4 - 788 \hat{s}^3 \hat{t} -
 1552 \hat{s}^2 \hat{t}^2 - 1208 \hat{s} \hat{t}^3 - 263 \hat{t}^4)\\\nonumber 
&+& m_{\psi}^{12} (1081 \hat{s}
^5 + 3875 \hat{s}^4 \hat{t} + 6582 \hat{s}^3 \hat{t}^2 + 7030 \hat{s}^2 \hat{t}^3 + 4537
 \hat{s} \hat{t}^4 + 1295 \hat{t}^5)\\\nonumber 
&+& m_{\psi}^{10} ( - 1296 \hat{s}^6 - 5283 \hat{s}^5 \hat{t}
- 10856 \hat{s}^4 \hat{t}^2 - 14170 \hat{s}^3 \hat{t}^3 - 12382 \hat{s}^2 \hat{t}^4 -
7039 \hat{s} \hat{t}^5 - 1958 \hat{t}^6) \\\nonumber
&+& m_{\psi}^8 (823 \hat{s}^7 + 2921 \hat{s}^6 \hat{t} +
 6357 \hat{s}^5 \hat{t}^2 + 11051 \hat{s}^4 \hat{t}^3 + 13065 \hat{s}^3 \hat{t}^4 + 9643
 \hat{s}^2 \hat{t}^5 + 4659 \hat{s} \hat{t}^6 + 1289 \hat{t}^7)\\\nonumber 
&+& m_{\psi}^6 ( - 224 \hat{s}^8
 - 615 \hat{s}^7 \hat{t} - 525 \hat{s}^6 \hat{t}^2 - 1609 \hat{s}^5 \hat{t}^3 - 4268 \hat{s}^4
 \hat{t}^4 - 4537 \hat{s}^3 \hat{t}^5 - 2267 \hat{s}^2 \hat{t}^6 - 843 \hat{s} \hat{t}^7 - 320
 \hat{t}^8)\\\nonumber 
&+&m_{\psi}^4\hat{s}\hat{t}(-36\hat{s}^7-727\hat{s}^6\hat{t}-1561\hat{s}^5\hat{t}
^2 - 894 \hat{s}^4 \hat{t}^3 + 290 \hat{s}^3 \hat{t}^4 + 117 \hat{s}^2 \hat{t}^5 - 389 
\hat{s} \hat{t}^6 - 192 \hat{t}^7)\\ 
&+&\left.3m_{\psi}^2 \hat{s}^3 \hat{t}^2 (44 \hat{s}^5 + 173 \hat{s}^4
 \hat{t} + 239 \hat{s}^3 \hat{t}^2 + 116 \hat{s}^2 \hat{t}^3 - 13 \hat{s} \hat{t}^4 - 19\hat{t}^5)\right\}/9.
\end{eqnarray}
In deriving (\ref{final}) we have used the following relation for the
long distance matrix element
\begin{equation}
<0|{\cal O}_1^{\psi}(^3S_1)|0>=\frac{3N_c}{2\pi}|R(0)|^2.
\end{equation}

\section{Numerical results and discussion}

\hspace{0.2in} For the purposes of numerical evaluation, we evaluate the ratio of
transverse momentum double scattering differential cross section to
corresponding single scattering contribution 
\begin{eqnarray}\nonumber
\rho^{(1)}&=&\left.\frac{d\sigma^D_{\gamma A}}{dQ^2_{\perp}}\right/
\frac{d\sigma^S_{\gamma A}}{dQ^2_{\perp}}\\
&=&\left.\int dy\frac{d\sigma^D_{\gamma A}}{dQ^2_{\perp}dy}\right/
\int dy\frac{d\sigma^S_{\gamma A}}{dQ^2_{\perp}dy}.
\end{eqnarray}
We perform the integration through rapidity $y$ in Eq.(\ref{leading}) and 
Eq.(\ref{final}) and the integration range are taken $|y|<0.4$. Before 
embarking on the numerical calculation, we have to know the non-perturbative parton
distribution $f_{g/A}(x)$ and multi-parton correlation function $M_g(x)$.

In Eq.(\ref{function}), the effective nuclear gluon distribution function
$f_{g/A}(x)$ should
have the same operator definion of the normal parton distribution with free
nucleon states replaced by the nuclear states. For a nucleus with atomic 
number $A$, we define
\begin{equation}\label{fga}
f_{g/A}(x)=Af_{g/N}(x),
\end{equation}
where $f_{g/N}(x)$ is normal gluon distributions in a free nucleon. In this 
effective nuclear parton distribution, we have neglected the EMC effect 
and the difference between the normal parton distribution in a free neutron 
and in a free proton. We adopt $f_{g/N}(x)$ from \cite{dis}.

The authors of \cite{luo,guo} proposed following phenomenological expressions
for the twist-four matrix elements that occur in soft rescattering:
\begin{equation}
M_i(x)=A^{1/3}\lambda^2f_{i/A}(x),
\end{equation}
where $i=q,\bar{q}$, and $g$. The $f_{i/A}(x)$ are the effective twist-2 
parton distributions in nucleus defined in Eq.(\ref{fga}). The constant $\lambda^2$ has units of mass
squared due to the dimension difference between twist-4 and twist-2 matrix
elements. The $\lambda^2$ should be determined by experimental measurement.
It was estimated in \cite{luo} by using the measured nuclear enhancement of
the momentum imbalance of two jets in photon-nucleus collisions, and was
found to be order of
$$
\lambda^2\approx 0.05-0.1GeV^2.
$$
In our following calculation we use $\lambda^2=0.05GeV^2$.

From the definion of the correlation functions in Eq.(\ref{distr}), the lack
of oscillation factors for both $z^-_2$ and $z^-_3$ integrals can in principle
give nuclear dependence proportional to $A^{2/3}$. The $A^{1/3}$ dependence
is a result of the assumption that the positions of two field strengths 
(at $z^-_2$ and $z^-_3$, respectively) are confined within one nucleon.

\begin{figure}
\begin{center}
\input{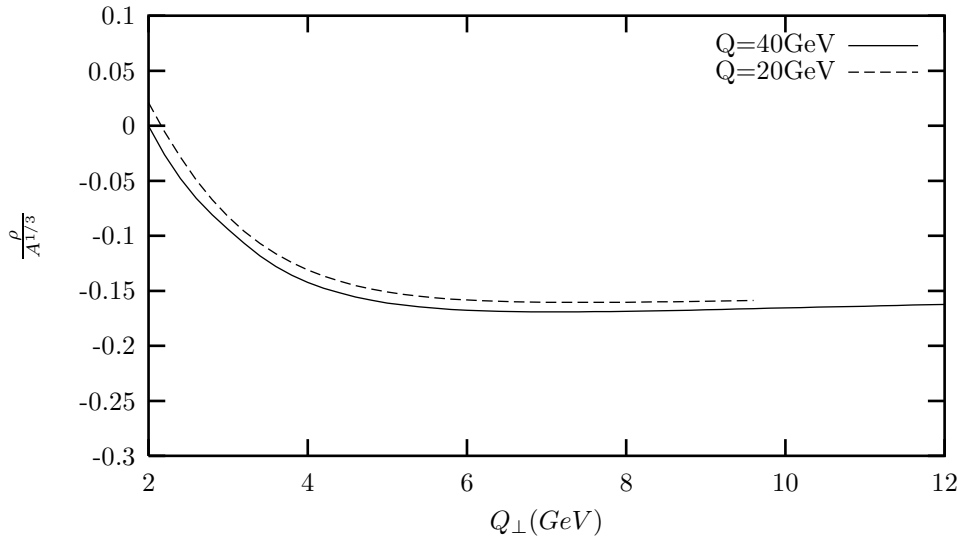}

\caption{The ratio of double scattering contribution and single
scattering contribution normalized by $A^{1/3}$ for $s=20GeV$ and
$s=40GeV$.}
\end{center}
\end{figure}

In Fig.9 we show the ratio of double scattering and single 
scattering versus $Q_{\perp}$. We plot $\rho^{(1)}$ as a function of $Q_{\perp}$ for
$s=(20GeV)^2$ and $s=(40GeV)^2$ respectively. The ratio is normalized by 
$A^{1/3}$. We find that the overall contribution of double scattering give 
negative sign, while it is positive in the processes of Drell-Yan and DIS. 
From Fig.9, it is obviously that the absolute value of the ratio increase 
with $Q_{\perp}$, this indicates that in high $Q_{\perp}$ $J/\psi$ is more
suppressed by the double scattering effect than that in low $Q_{\perp}$. We
find that the two curves for $s=(20GeV)^2$ and $s=(40GeV)^2$ are very 
close, this indicates that at this range the ratio is
almost independent of the invariant $s$.
Although our quantity result is not large, about $17\%$ at $Q_{\perp}
=8GeV$ for $s=40GeV$, but if we include the extra $A^{1/3}$ factor, it can
give observable effect in large $A$ case. More over our result lies in 
reducing some theoretical uncertainties.
 
In summary, we have applied LQS method, which was developed in inclusive DIS
process, to study the multi-scattering effect on $J/\psi$ 
photoproduction. We consider the interaction between nuclear matter and
$c\bar{c}$ pair produced from the hard collision of a photon
and a parton in nucleus. This interaction undergoes through either quark or antiquark
absorbing a soft gluon. We find the lowest order double scattering has not a
great influence and only contributes
a slight suppression comparing with the single scattering cross section.
Our analysis can be extended to study the multi-scattering
effects in $J/\psi$ production from hard-nucleus or nucleus-nucleus 
collisions, which will involve initial state strong interactions that are
absent in the case of photoproduction. These two hadron production processes 
will be more complicated because for initial state we must consider two types 
of double scattering, soft-hard scattering and double hard scattering. 
 
\centerline{\bf Acknowledgment}

H.W.Huang would like to thank Prof. H.Satz for useful discussions, and the
referee for helpful suggestions which improved the original version of
this paper. This work was supported by the Alexander von Humboldt Foundation.

\vfill\eject
\end{document}